# APPLIED DESIGN THINKING IN URBAN AIR MOBILITY: CREATING THE AIRTAXI CABIN DESIGN OF THE FUTURE FROM A USER PERSPECTIVE

F. Reimer (https://orcid.org/0000-0001-6666-6903),
J. Herzig
L. Winkler
J. Biedermann
F. Meller
B. Nagel

DLR Institute of System Architectures in Aeronautics,
German Aerospace Center
Hein-Saß-Weg 22, 21129 Hamburg, Germany

Contact: fabian.reimer@dlr.de

## Abstract

Design thinking is essential for user-centered cabin design concepts in future transportation vehicles, as it facilitates the identification of user needs, creative problem-solving and iterative development to ensure optimal user experiences and satisfaction. In the exploration of future air taxi cabins , user acceptance is widely recognized as a critical factor. To ensure a high level of acceptance for such concepts, the direct involvement of potential user groups in the early design process through user-centered design approaches, offers a highly effective solution to provide a time efficient and requirement-based concept development process for novel transportation concepts.

In the course of developing digital and future aviation cabin concepts at the German Aerospace Center, the exploration of user-centered and acceptance-enhancing methods plays a central role. The challenge here is to identify the flexible range of requirements of different user groups for a previously non-existent transport concept, to translate these into a concept and to generate a rapid evaluation process by the user groups. Therefore, this paper aims to demonstrate the application of the user-centered Design Thinking method in the design of cabin for future air taxis. Based on the Design Thinking approach and its iterative process steps, the direct implementation is described on the combined airport shuttle and intracity UAM concept. The main focus is on the identification of key user requirements by means of a focus group study and the evaluation of initial cabin designs and key ideas by means of an online survey. Consequently, the creative design process of a digital prototype will be presented. In addition to an increased awareness and acceptance among the population towards a novel mode of transportation, the application of the Design Thinking methodology offers a flexible and user-centered approach for further testing and simulation scenarios.

## NOMENCLATURE

| | |
|---|---|
| 3D | Three Dimensional |
| DLR | Deutsches Zentrum für Luft- und Raumfahrt |
| DT | Design Thinking |
| eVTOL | Electric takeoff and landing |
| Kg | Kilogram |
| Km | Kilometer |
| Km/h | Kilometer per hour |
| M | Meter |
| MR | Mixed Reality |
| RPAS | Remotely Piloted Aircraft System |
| UAM | Urban Air Mobility |
| UAV | Unmanned Aerial Vehicle |
| VR | Virtual Reality |

## 1. INTRODUCTION

Urban air mobility (UAM) is an emerging transportation concept that involves the use of aerial vehicles (air taxis) in urban environments. While UAM has the potential to revolutionize the way people and goods are transported, its successful implementation depends on public acceptance. As the interface between air taxi and the passengers, the cabin and the interior design play a crucial role in shaping passenger's perception and the attitude towards the safety, comfort, usability, flight experience and might have a substantial impact on people's acceptance towards future air taxis. However, as transportation concepts like air taxis have not existed before, the low range of user experiences and the identification of potential requirements and needs are particularly challenging.

Despite the advanced technological development of air taxis and expected flights in the coming years by Volocopter [1]., the German Aerospace Center (DLR) has conducted various studies to explore the question of societal acceptance

Consequently, it can be displayed, that the public acceptance plays a crucial role in the development and realization of drones in the future as stated by Eißfeld et. al., investigating the acceptance of Civil Drones in Germany in 2020 [2].

As part of a study involving 832 participants, the societal acceptance towards drones was examined. While 38% of the subjects had a predominantly negative attitude towards drones, 53% demonstrated a predominantly positive attitude.

The disagreement and rather sceptical attitude among the general population towards air taxis were further demonstrated in a subsequent study by End et al. [3]. In this study, 19% of the participants exhibited a very negative attitude, while 8% held a positive stance.

In addition, the literature highlights the factors of safety and privacy as particular significant factors contributing to the sceptical attitude of the public towards air taxis [4], [5], [6] .

Another reason for the potential rejection of air taxis, as stated by the German Aerospace Industries Association (BDLI). Within an acceptance study about drones and airtaxis in 2022 and 2.055 testsubjects in Germany, the limited information and knowledge about air taxis among the general population was discovered to be a major factor [7].

This shows parallels to the literature review by Kellermann & Fischer (2020), where the provision of information as well as the transparency towards the public play an essential role in terms of an increased level of acceptance [8].

In terms of providing a high level of acceptance of future innovations by the general public, user-centered methods such as Design Thinking have proven to be particulary effective. n the development of user-centered innovations, the Design Thinking methodology has demonstrated its effectiveness. It was initially developed by Professors Kelley and Winograd at Stanford University and later established in Europe by co-founder Hasso Plattner from 2007 Design Thinking is an approach that emphasizes comprehensive understanding and deep, empathetic insight into user behaviours, desires, fears, and environments [9]. According to Meinel et. al., Design Thinking combines a focus on the end user with various multidisciplinary approaches and iterative and continuous optimization [10].

Therefore, it is crucial to understand users' needs and integrate them into the early design process of future airtaxi cabin concepts to ensure a high level of acceptance for this new transportation mode. At the German Aerospace Center (DLR), aeronautical cabin research is focused on developing and digitally assessing cabin systems and designs for future user groups and their needs. The DLR project Horizon UAM aims to develop a holistic, digital, and user-centered UAM concept. Following the "Inside Out" approach, the comprehensive development and description of the UAM system are based on cabin design. In addition to technical considerations, certification requirements, and infrastructure factors, acceptance, the development of appropriate user testing scenarios, and the inclusion of different user groups in the digital development process play a central role.

This paper intends to close the gap between user and designers adapting the user-centered design method Design Thinking on the cabin design approach for future Airtaxi cabin design concepts.

A central focus lies in the step-by-step description and implementation of the methodological pillars of the classic Design Thinking approach in the cabin design process of future air taxi cabins. The initial process steps prioritize the identification of key requirements, fears, needs, and experiences of potential user groups. Another focus is on describing the conceptual design outcomes and the parallel evaluation process involving user groups. The goal of these endeavors is to present an applied and methodological framework for user-centered design of future air taxi cabins. This involves emphasizing conceptual development while actively involving the population in the development process of this novel mode of transportation to enhance transparency and acceptance.

## 2. FUNDAMENTALS

This section provides a description of the Design Thinking methodology in the context of Human Centered Design, along with the general definition of the six characteristic process steps and terminologies as the foundation for practical application in the design process of the UAM cabin concept. Furthermore, it positions the conceptual development of a user-centered cabin concept within the background of the DLR project Horizon UAM and the overall system and vehicle development of an air taxi concept.

### 2.1. Design Thinking Method

Design Thinking (DT) is a human-centered design approach aimed at generating innovative concepts based on a deep understanding of human needs [11].

Within the spectrum of human centered design methods, Design Thinking offers a creative and effective approach that emphasizes users' emotions, enabling more effective solutions for various stakeholder requirements [12].

The high potential of this method for developing complex systems is visible by the increasing number of scientific publications related to Design Thinking [13].

Fundamentally, Design Thinking follows a user-centered, empathetic, and analytical approach to solving complex problems. However, since there are no strictly defined or fixed process steps, different interpretations of the process steps can be observed in industry and research. IBM for example follows the DT approach called "The Loop," which emphasizes an iterative and continuous cycle consisting of the process steps of observing, reflecting, and making (FIG 1) [14]. All three steps cover on particular part within an infinite loop, which displays the continuous design and and feedback process throughout a new design.

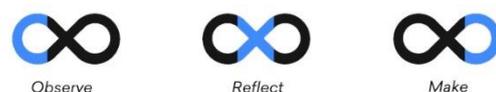

**FIG 1**  IBM Design Thinking process "The Loop" [15]

A further approach is being provided by the so called „Double Diamond" process (FIG 2). The official Double Diamond design model has four stages: **Discovery, Definition, Development and Delivery** and incorporates a divergent and a convergent design stage [16].

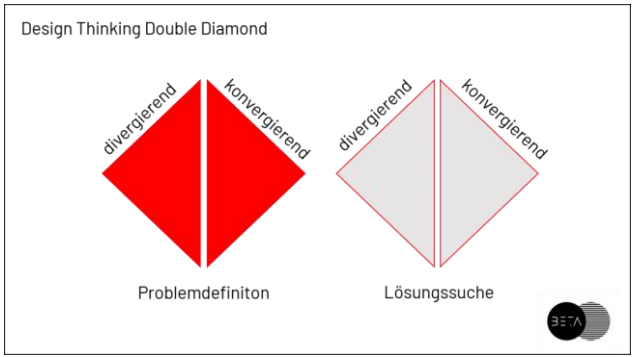

**FIG 2**    Design Thinking "Double Diamond" [17]

Nevertheless, the most well-known approach is the Stanford d.school Design Thinking approach with its six steps ´Empathize´, ´Define´, ´Ideate´, ´Prototype´, ´Test´ and ´Assess´ (FIG 3).

## Design Thinking Process Diagram*

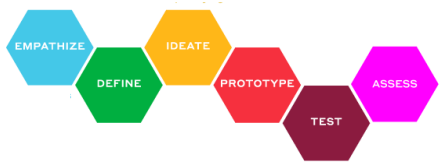

**FIG 3**    Design Thinking model proposed by the Hasso-Plattner Institute of Design at Stanford (d.school) [18]

In Design Thinking, the individual phases and the iterative process between them play a crucial role. In the empathy phase, the focus is on observing and understanding the users, capturing their habits and behaviors in their environment. In the Define phase, the central problems of the user are filtered and defined collaboratively. In the Ideate phase, different ideas and radical designs are generated. Quantity is emphasized over quality, allowing for a wide range of ideas for initial decision-making. In the prototype and test phase, selected ideas are typically created as rapid prototypes and tested, while always keeping a loop back to the initial problem statement. In the evaluation phase, users and other participants can provide feedback as experts or recipients, assisting in the assessment process. Key characteristics include a free and creative approach and the constant iterative testing and evaluation of ideas following the "Fail-Fast" principle [19].

This variant, with its six process steps, forms the methodological foundation for the user-centered and iterative design process of future air taxi cabin concepts within the scope of this paper.

## 2.2. Horizon UAM Project and Operation Scenarios

Since 2020, the German Aerospace Center (DLR) has been conducting research on Urban Air Mobility as part of the Horizon UAM project, focusing on factors such as efficiency, safety, feasibility, sustainability, and more (FIG 4). All areas are part of the overall air taxi system and have an influence on the acceptance of the population.In collaboration with ten DLR institutes and external partners like NASA and Bauhaus Luftfahrt, the collaborative research of these subsystems with a focus on acceptance is the essential core of DLR's exploration and research of future air taxi technologies.

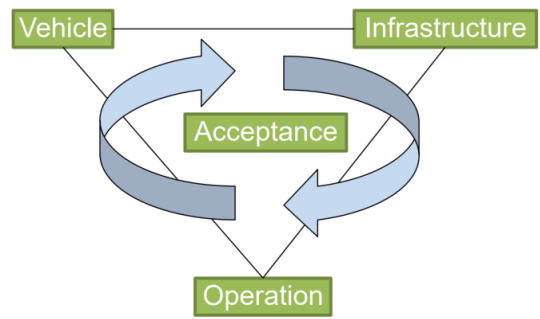

**FIG 4**    DLR project Horizon UAM framework [20]

As a starting point for the concept development, various UAM use cases and technology scenarios were designed, including Intra-City, Mega-City, Airport Shuttle, Suburban-Commuter, and Intercity. The characteristics of the scenarios were defined in the course of a joint workshop and on the basis of technically feasible parameters.The main focus for the present work is the combination of the following use cases:

**Intra-City Scenario:**

- Transport range: up to 50 km
- Speed: up to 100 km/h
- Seats: up to four

**Airportshuttle Scenario**

- Transport range: up to 30 km
- Speed: up to 150 km/h
- Seats: up to four

To develop a collaborative concept based on the Design Thinking method, a combination of both use cases was chosen to create a user-centered cabin design for a multifunctional and versatile short-range air taxi concept (FIG 5). The figure shows a graphical representation of different and self-contained scenarios that could be connected in perspective.

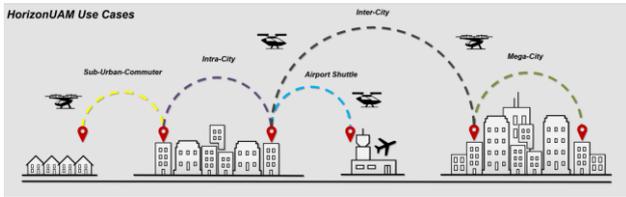

**FIG 5**    Horizon UAM Use Cases [21]

## 3. METHODICAL DEVELOPMENT OF AN INTRACITY/AIRPORTSHUTTLE CABIN CONCEPT

The following section describes the particular process steps of Design Thinking (Stanford d.school) in course of the design process of an airport shuttle and intracity air taxi cabin concept. Based on the characteristic steps of Design Thinking, the main outcomes of the design process are listed and explained.

### 3.1.  Empathy

In the empathy phase, two factors are particularly important. Firstly, to create an overview of the current state of the art in interior design of air taxis, and secondly, to understand and observe participants in a focus group study. Alongside the state-of-the-art analysis, the execution of the focus group study is described.

**State of the Art:**

Similar to the research on air taxis, autonomous transportation of people plays a significant role in the automotive industry. Unlike in air taxi research, there are already advanced innovations in the field of cabin design here and thus provide a suitable basis for initial analyses of the state of the art. For instance, innovative models often feature strong color contrasts, such as dark brown/grey tones for a luxurious look and white/cream tones for cleanliness and high quality. Green tones and wood optics are used to convey durability and environmental consciousness. Bionic forms in storage compartments or ceiling columns imitate nature, emphasizing the connection to sustainable design. Large windows provide passengers with a clear view of their surroundings, fostering a connection to nature. First impressions are crucial in automotive innovation, leading to unique door concepts that aim to captivate potential customers. Vehicle designs serve as essential indicators for feasible innovations in the UAM domain. Leveraging the familiarity of automobile design is important for creating recognition value and a sense of familiarity and security [22].

Research on the existing UAM vehicles shows that this branch has learned from the innovation in the automotive sector. A lot of the companies developing a UAM vehicle is still in the early stages of the design process and has not revealed the cabin concepts yet. Those who have, show similarities with design approaches of most modern cars. The interiors are based on strong colour contrasts and minimalistic design, conveying a sense of connection to the automotive sector. Clarity in the design is here as well achieved through bionic window shapes and large windows, meant to enhance the flight experience. Seats are most often arranged according the automotive standard, creating recognition value. An exemplary comparison between a modern car interior design (Moia taxi) and the Lilium eVTOL cabin design is shown in Figure 6.

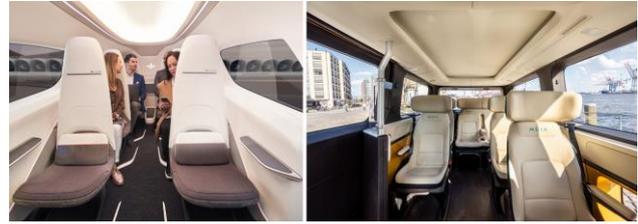

**FIG 6**    Volocopter (left) [23] and Moia (right) [24] interior examples

In course of the state of the art process, different future concepts of the automotive and eVTOL industry have been investigated to collect a basis overview for the following design process steps.

**Focus Group Study on passenger requirements for future air taxis**

In December 2020, a focus group study was conducted to gain a fundamental understanding of potential user groups' attitudes towards UAM and gather information about the spectrum of requirements for cabin design [25]. Four focus groups have been investigated in December 2020 with a total of 16 participants from Germany. The studies were divided into three parts, and the execution of each step is described below. Additionally, a summary of the key findings for the conceptual development of a cabin concept within the Design Thinking approach is provided.

- Part 1: Public Transport Preferences

In the first part of the focus group study, participants were asked about their preferences for public transportation. Alongside naming their favorite modes of transportation, participants were also requested to identify both positive and negative aspects. This part was conducted to gather individual experiential insights from public transportation and raise participants' awareness of air taxis as a transportation mode.

- Part 2: Public Transport Interior Preferences

In the following step, the participants were asked about their general preferences regarding the interior of public transportation. The results were recorded and divided into four levels according to Sorrel Brown's importance Likert scale [26]. The results can be seen in the following table (TAB 1).

| Requirements for Cabin Interior | | | |
|---|---|---|---|
| **Very Important Req.** | **Important Req.** | **Moderately important Req.** | **Slightly important Req.** |
| -Existance of windows<br>- Large panorama windows preferred<br>- Good Air conditioning<br>- Climate not being too warm or too cold<br>- Temperature adjustment (also to prevent bad smells)<br>- Bacteria filtering technology (cleanliness and hygienic aspects)<br>- Quiet Air conditioning<br>- Small noise level e.g. with noise cancelling headphones<br>- Presence of electric sockets for charging electronic devices | -Separate compartments for privacy, working, resting, sleeping<br>- Sufficient legroom<br>- Good lighting<br>- Proper working light<br>- Bright environment, particularly with high ceilings<br>- Entering and leaving the vehicle should be comfortable<br>- Wide aisles for easy moving through vehicle<br>- Entries and exits on different sides | - Seat comfort: Not too hard, not too soft seat (no hard materials like wood)<br>- Seats should be adjustable<br>- To mentain privacy, distance between seats in y-axis<br>- On lengthier journeys area to move around<br>- Place to store notebooks, food and drinks (tables preferred)<br>- On Board entertainment system needed for movies and audiobooks<br>- Passenger information, travel infos (speed, height, temperature etc.), connection infos to other vehicles<br>- High level of cleanliness<br>- On board toilets | - High seats, so that legs are not overly bowed<br>- Seat warmers<br>- Sufficient headroom<br>- Muted colours, a suitable number of waste bins<br>- Nice floor temperature and safe/non-slippery floor<br>- Provision of meals during journey |

**TAB 1**  Cabin Interior Preferences (Credits:DLR)

Requirements with high or very high priority include comfort-related parameters such as seat comfort, thermal comfort, and low noise levels. Easy-to-understand operation, individual comfort settings, and the option to adjust privacy levels were also emphasized.

Less prioritized were aviation-specific features like in-flight dining, toilets, or on-board entertainment. Availability of laptop storage, seat heating, or pleasant floor temperature were also considered less important [27].

The task was conducted using the collaborative platform Mural and the communication platform Skype for Business. An exemplary board with solution approaches is shown in the following figure (FIG 7).

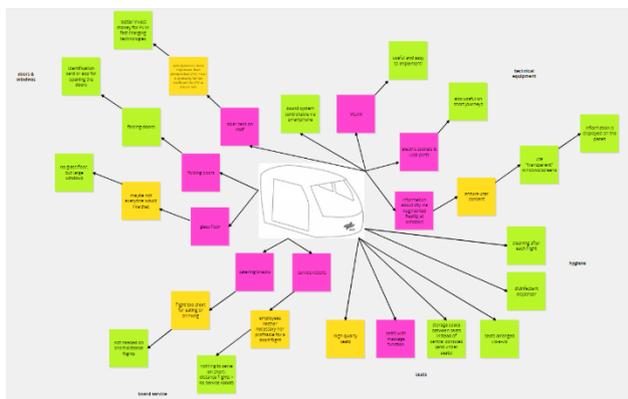

**FIG 7**  Exemplary Overview: Red (Dreamer), Yellow (Critics), Green (Realist); (Credits:DLR)

Based on the first two parts and the collaborative cabin design process, various solution approaches were developed, leading to four main areas of focus for the requirements: Comfort & Experience, Safety & Security, Luggage Storage, and Seating & Configuration.

## 3.2.  Define phase

In the Define phase, the focus was set on summarizing and organizing the insights gained previously. Typically, the central task was to consolidate key requirements using personas. Additionally, and as part of a complete air taxi system, the definition of architectural space parameters and central weight parameters played a crucial role.

**Persona Definition**

To make the diverse range of requirements from different user groups more tangible, personas were developed within the framework of the Design Thinking method. Personas serve as representative profiles for a broad range of users, facilitating a more targeted design process based on defined requirements [28]. To cover a wide range of age groups, three personas were created.

•      Greta Hermann, female, 62 years old, small town, teacher, generation „Boomer"

•      Tim Klaussen, male, 35 years old, suburban, consulter, generation „Y"

•      Clara Meyer, female, 19 years old, metropolitan, student, generation „Z"

The exemplary definition of the persona profile for "Clara Meyer" can be seen in the following figure (FIG 8). In addition to demographic data on the right, persona characteristics related to travel behavior (bottom), biographical data (middle, top), and character traits (right). In addition, collected key requirements for air cabs are bundled on the basis of this persona (middle, bottom).

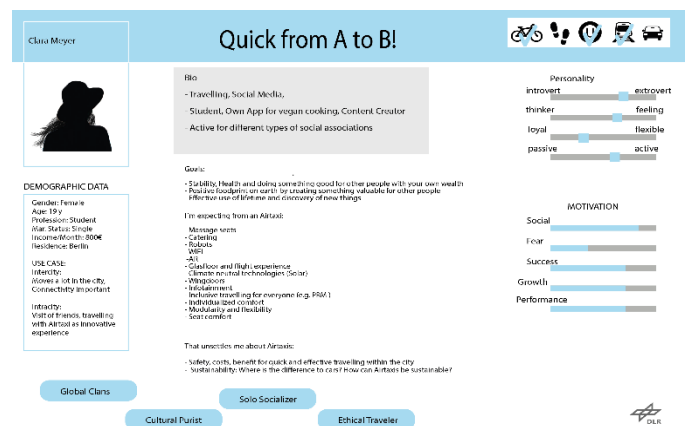

**FIG 8**  Exemplary Persona Definition; (Credits:DLR)

**Cabin Layout Definition & Overall Parameters**

Within the definition of a cabin design, ergonomic considerations play an important role. For the integration process of the cabin into the overall vehicle system, central requirement parameters have been defined as a first baseline:

•  Payload weight: 90 kg (+20 kg optional and additional luggage weight)

•  Piloted vehicle with option for autonomous flight option

•  Four Seats

•  Usability for PRM (storage of wheelchair)

For the detailed design and definition of an initial cabin layout, the previously defined requirements, as well as ergonomic and anthropometric standards, were considered. Comfort parameters related to seat spacing and width were derived from common dimensions found in business class parameters in commercial aviation. The necessary storage space was defined based on common carry-on luggage dimensions travel suitcase measurements and dimensions for standard wheelchairs.

Basically, a layout was designed that meets the basic anthropometric requirements of different passengers with the smallest and largest possible physical dimensions. Due to the use case airport shuttle it was determined that at least four pieces of luggage should be carried in the cabin. In addition, a seat layout with two rows and two seats per row have been specified. The described layout can be seen in the following figure (FIG 9). The figure shows the schematic representation of the central and anthropometric limit parameters for the different areas of the cabin.

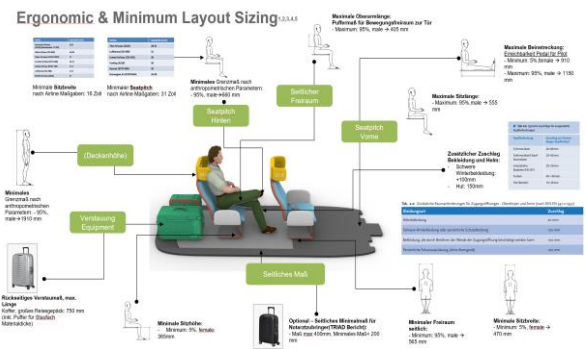

**FIG 9** Cabin sizing based on ergonomic parameters; (Credits:DLR)

A combination of ergonomic requirements of different passengers' types and technical requirements of the overall vehicle and system development process formed a final layout with defined dimensions for further design processes. Due to the use case airport shuttle and up to four potential passengers it was determined that at least four pieces of luggage should be carried in the cabin. In addition, a seat layout with two rows with two seats per row was specified. The described layout can be seen in the following figure (FIG 10).

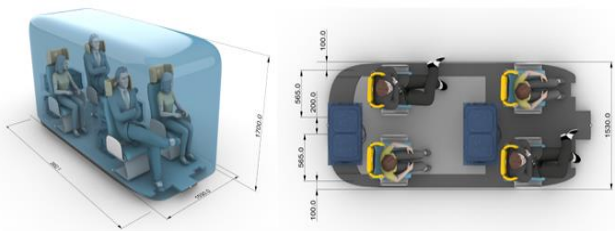

**FIG 10** Final and usable space for cabin integration; (Credits:DLR)

### 3.3. Ideation phase

**Idea development process**

In the Ideate phase, various ideas and concept focuses were developed based on the defined constraints from the Empathize and Define phases. In addition to different group seating scenarios, ideas for customizable areas through the strategic use of partitions were designed. Seat designs with recognizable characteristics from the automotive and aviation sectors were also created as exemplary ideas for the seating arrangement in a future air taxi cabin scenario. Given the significant role of the Airport Shuttle and Intracity use case, different options for accommodating large luggage were developed. The thematic breakdown and overview of the design ideas can be seen in the following diagram (FIG 11).

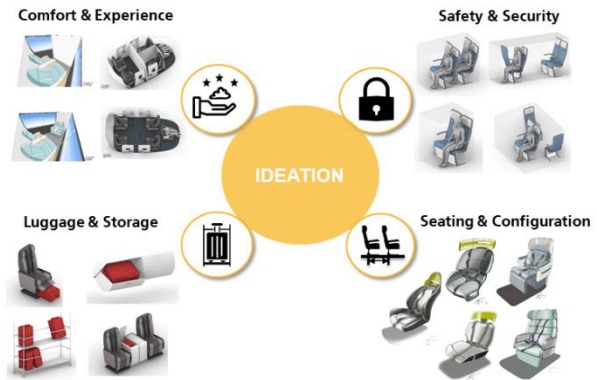

**FIG 11** Ideation breakdown overview; (Credits:DLR)

**Onlinesurvey: Definition of air taxi cabin design parameters and the evaluation of design ideas.**

The process of ideation, involving potential user groups and central requirements for overall system dimensioning, is complex. To involve as many potential user groups as possible directly in the ideation and decision-making process, an online survey was conducted in Germany in 2021.

The onlinesurvey resulted in 202 valid datasets of participants from various demographic groups in Germany. Within the study, six UAM cabin designs were presented and evaluated [29].

Within the context of the previously defined priority areas in the Focus Group Study (see chapter 3.1), idea approaches related to Safety & Security, Comfort & Experience, Luggage, and Seating & Configuration were evaluated. With regard to the defined short-haul use case, participants were presented with a scenario of a fully occupied air taxi with a travel time of ten to 15 minutes.

The key findings of this study are described as follows:

• Seating & Configuration:

In the first part of the study, eight different ideas for the arrangement of a four-seat configuration were presented. In the front area, a pilot seat was placed, which was not included in the seat selection evaluation. The respondents were asked to rate the concepts based on two perspectives: First, from the perspective of a flight without an accompanying person and then from the perspective of traveling with an accompanying person. Responses were given using a single-choice format with six options: 'very bad', 'bad', 'neutral', 'good', 'very good', and 'no answer'.

The key results are depicted in the following figure (FIG 12).

**FIG 12**   Results Seating & Configuration (Credits:DLR)

Overall, it was demonstrated that communication and proximity with eye contact play an important role in the scenario of traveling with an accompanying person. In the scenario without an accompanying person, configurations with direct eye contact with strangers or significant distance from accompanying persons were rated the least favorably.

• Safety & Security:

Based on the previously described focus group study, it became evident that travelers value protection from fellow passengers and privacy during the flight. In the second part of this study, participants were presented with six different ideas for physical separation from fellow passengers for evaluation. The scenario and evaluation criteria were consistent with the "Seating & Configuration" scenario assessment (FIG 13).

**FIG 13**   Results Safety & Security (Credits:DLR)

In the scenario with an accompanying person, it became evident once again that communication and proximity to the accompanying person in the adjacent seat play an important role. Complete separation through partitions from fellow passengers was rejected in this scenario. In the scenario without an accompanying person, separate areas with a stranger and direct eye contact were strongly rejected. Isolation from the rest of the cabin, constant eye contact, and sharing a common foot space with a stranger were cited as key reasons for the negative ratings.

• Comfort & Experience:

To assess different seating ideas and the associated seating comfort, five different seat concepts inspired by known concepts from the aviation industry (business class seat, first-class seat), the automotive industry, as well as novel seat concepts were presented for evaluation by the participants. The key findings are illustrated in the following figure (FIG 14).

**FIG 14**   Results Comfort & Experience (Credits:DLR)

The seats inspired by higher-class airline seats were particularly positively rated. Reasons for this included expected seating comfort, presence of armrests, and especially the U-shaped modern headrests for increased privacy. Less familiar seat models with novel shapes and no arm rests were negatively rated.

• Luggage & Storage:

For the evaluation of different luggage storage concepts, seven different approaches were presented. Participants could choose between the response options ´impractical`, ´rather impractical`, ´partly practical`, ´rather practical`, ´practical`, and ´no answer`. The approaches with the highest approval are shown in the following figure (FIG 15).

**FIG 15**   Results Luggage & Storage (Credits:DLR)

Key factors for a positive rating included sufficient storage space, easy accessibility of luggage during the journey, individual storage options, and secure attachment of the luggage.

## 3.4.   Prototype phase

Prototyping is an experimental process used to create digital or physical prototypes based on the insights gained from previous stages. The following section illustrates and describes the creation of a prototype using digital sketches including 3D designs within the Design Thinking process step ´Prototype`. The results obtained from this stage form the basis for the final evaluation process. The prototyping process is described below, based on the previously defined priority areas.

**Seating & Configuration**

In response to the high demand and user preference for a more traditional layout, a configuration with two seats per row positioned in the direction of flight was chosen (FIG 16).

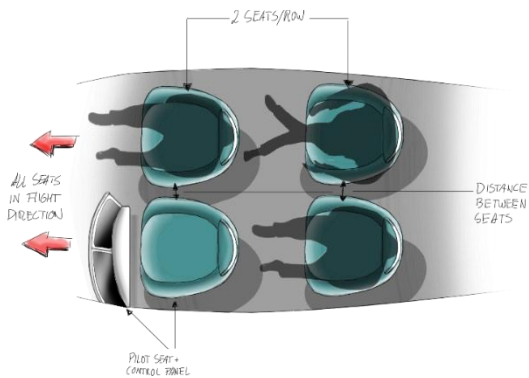

**FIG 16** Conceptual Seat Layout (Credits:DLR)

In addition to the adequate seat pitch, seat width and additional space between the seats, a cockpit was also integrated in the front-left area. In the course of the overall concept within the main work package, three passengers and one pilot were previously defined as concept payload. The detailed design of the cockpit was not the main focus here, which is why only a control panel was integrated as a placeholder. Accordingly, possible requirements with regard to protective devices, control modules and screens were not considered.

**Safety & Security**

Looking at the acceptance aspect of future air taxis, the area of safety and security plays a fundamental role [3]. Based on the user requirements, numerous concerns and wishes were expressed to ensure protection from fellow passengers (e.g. violence by fellow passengers) and individualizable protection concepts for increased privacy when traveling with strangers. To ensure this, three protection concepts were developed for the overall concept.

A first approach is shown by the novel positioning of the seats based on a rotated arrangement is shown below by means of an example (FIG 17).

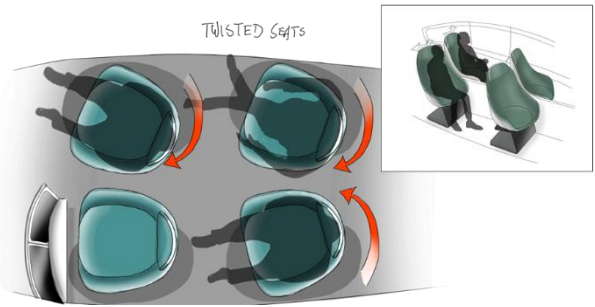

**FIG 17**: Example of rotated seat idea (Credits:DLR)

The novel position offers various advantages for passengers. On the one hand, the decision for this seating position was made to differentiate from previous and familiar seat layouts and to convey the novelty and modernity of this means of transportation. On the other hand, the novel seating position guides passengers to look out the window during the flight and experience the flight more and ensures an easier boarding and deboarding process. At the same time, this position necessarily provides additional distance from fellow passengers, which can lead to increased levels of safety and privacy.

Another protection option is offered by the U-shaped headrest, which is integrated into all seats. This form of headrest is currently already being used in a variety of cabin seats or public transportation systems to create a sense of spatial, visual, and acoustic separation from fellow passengers in the simplest way possible. The following figure (FIG 18) shows a first draft for this concept.

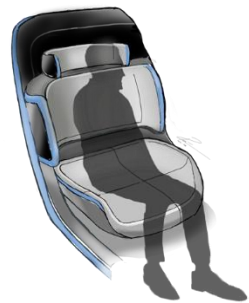

**FIG 18** Headrest and seat idea (Credits:DLR)

As a further protection and separation option, a central separation module was accessed between the rows of seats (FIG 19). This type of separation is already used in the automotive sector and serves as an additional separation option between passengers in the course of the air cab concept. In addition, the use of this separation module creates an individual area for each passenger, which could have a positive influence on the individual perception of comfort.

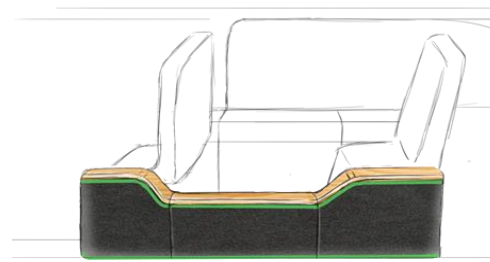

**FIG 19** Separation module (Credits:DLR)

**Comfort & Experience**

In the course of the user research phase, the area of comfort & experience was identified as a key requirement and plays a decisive role in the acceptance of future air taxi cabin concepts. The following figure (FIG 20) shows an idea sketch for the area of seat comfort.

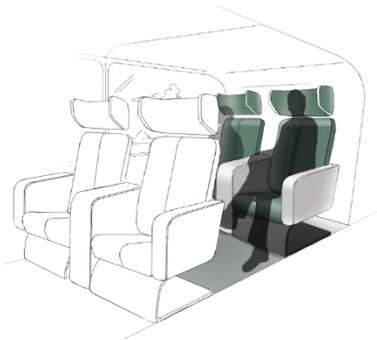

**FIG 20**  Comfort seat arrangement example (Credits:DLR)

Since passengers will spend most of their travel time in a seat during flight, the seat is a key element and a main interface between the user and the overall cabin. As a reference for the design of the seat position, a minimum seat pitch of 31 inches was defined as a basis here. A minimum seat pitch of 17 inches was defined for the seat width. The dimensions were taken from common positioning dimensions of aircraft cabins with increased comfort standards. To ensure sufficient headroom at a cabin height of 1.60m, no storage compartments or displays were placed above the seating areas. In the detailed design of the passenger seats, a separate armrest and headrest for increased privacy and optimum seating comfort also play a special role. The following figure (FIG 21) shows an idea sketch for the positioning and storage of folded standard wheelchairs.

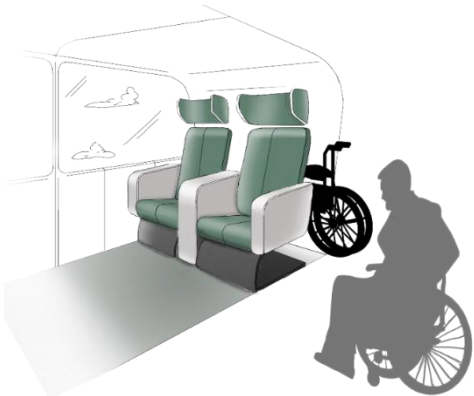

**FIG 21**  Wheelchair accessanbility idea (Credits:DLR)

To ensure a barrier-free cabin design, it must be possible to carry one's own wheelchair. For the overall concept, the storage space in the rear area was defined first. With the help of a ground crew and/or a pilot, the wheelchair could be stowed in the rear area of an air taxi and reached via a side flap from outside of the vehicle.

Another concept focus has been set on the flight experience (FIG 22). The provision of a positive passenger experience plays a special role when it comes to acceptance of new types of transportation and must be transferred precisely to the requirements, especially in cabin design [5]. In the course of the focus group study and the survey, the desire for a good view was particularly emphasized, so that a good view for all passengers must be guaranteed for the sub concept.

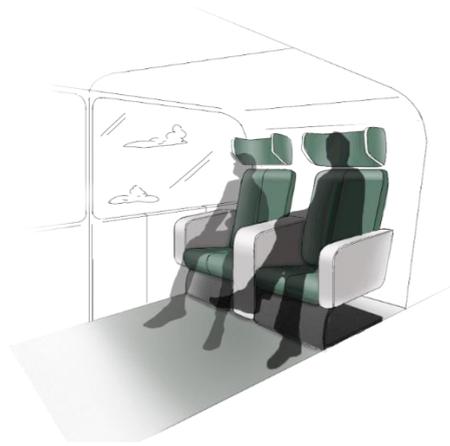

**FIG 22**  Outside view and experience (Credits:DLR)

A further aspect for the flight experience is a good balance between privacy, safety and group travel options including a possibility to share this new experience with other passengers. To create a basic distance between passengers, the division of the cabin into two rows of seats with two seats already provides a basic level of distance to other passengers. Depending on the individual wishes (group travel or singles travel), seats in one row or one behind the other can be selected individually. In the respective seat rows, the armrest as well as the headrests serve as elements that automatically create a distance between the passengers.

**Luggage**

Baggage stowage was defined as a central comfort parameter for user groups and can have a decisive influence on the travel experience and comfort. Since accessibility during the flight was named as an important aspect, the sub concepts could already be designed for stowage concepts in front of the passenger seats when defining the ideas. The four sub-concepts are shown and described below.

• Luggage Sub Concept 1 (Fixation Bars):

In this concept, curved bars with two-point fixation were integrated in the floor in front of each seat. The shape of the bend here makes it possible to place standard travel suitcases. This simple holding device makes it possible to fix the suitcase simplified and directly in front of the seat (FIG 23).

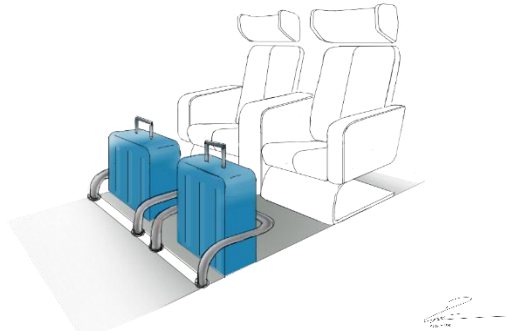

**FIG 23**  Luggage Sub Concept 1 (Credits:DLR)

- Luggage Sub Concept 2: Fixation Belt

Additionally, suitcases can be fixed with the help of a belt mechanism (FIG 24). Since the mechanism might already be familiar to most passengers from the automotive sector, this mechanism is particularly promising due to its simple and intuitive operability. In addition, the belt function offers a particularly simple and intuitive option for fixing folded standard wheelchairs or strollers.

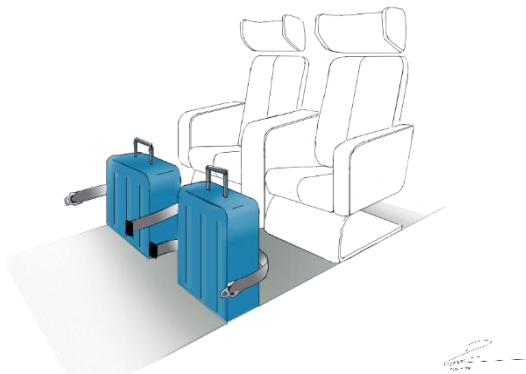

**FIG 24**  Luggage Sub Concept 2 (Credits:DLR)

- Luggage Sub Concept 3: Protection Option

Due to the expected and frequent passenger changes in the course of the short-haul vehicle operation, a high degree of wear and tear of components caused by suitcases during loading is to be expected (FIG 25). For this reason, an impact protection device was designed as a sub concept.

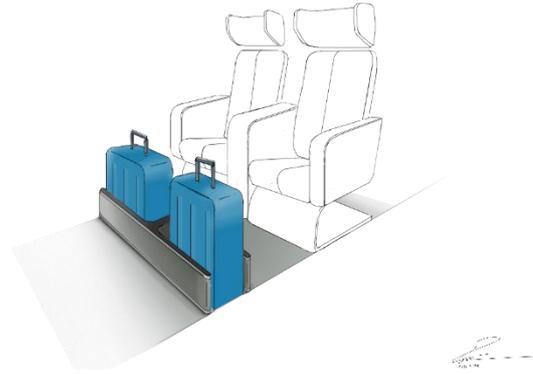

**FIG 25**  Luggage Sub Concept 3 (Credits:DLR)

### Integrated and final concept

Figure 26 shows the overall view of the integrated concept in the 3D model. In the overall concept, three seats have been rotated in the direction of the windows, while the seat in the front part on the right-hand side remains unchanged. In this scenario, a pilot is included, but this concept should already be usable for autonomous operation cabin concept designs. Accordingly, the installation space for the pilot seat was chosen to provide an easily adaptable seating scenario. The cockpit has not been further detailed in the course of the development, but a modular control panel was added as a control unit.
In addition to the seat's rotatability, the u-shaped headrest is a distinctive component of the concept. The rotating headrest allows passengers to determine their own level of privacy and. Finally, this creates further options for an increased level of privacy. A central module between the seats is used to store smaller items or charge cell phones via a small surface on top.

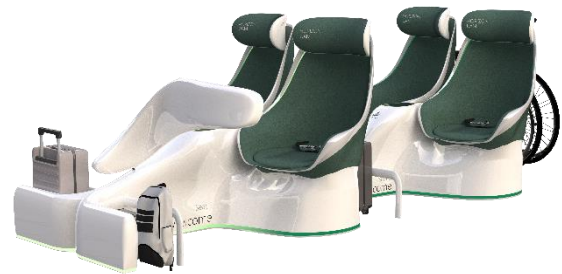

**FIG 26**  Perspective view final UAM Cabin Design Concept (Credits:DLR)

The perspective front and rear view clearly shows, that a simple stowage concept was chosen. Closable compartments and storage options were deliberately disregarded in order to prevent hand luggage and personal devices from being forgotten in the short-haul concept. A combination of Sub Concept 1 and 3 was chosen for the hand luggage holder (see 3.3).

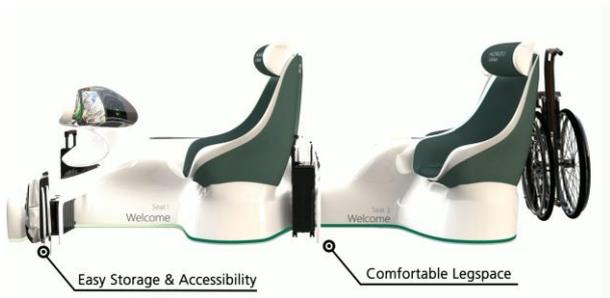

**FIG 27** Side view (Credits:DLR)

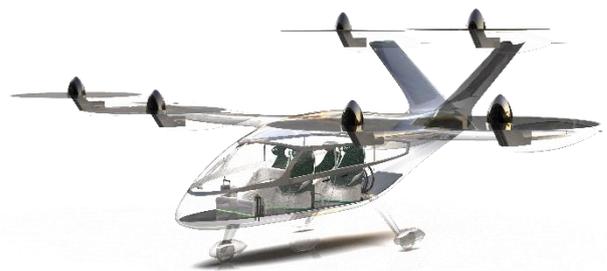

**FIG 29** Final cabin concept integrated in DLR Tiltrotor concept (Credits:DLR)

This ensures an easy stowage and removal of luggage (FIG 27). At the same time, a simplified accessibility during the flight is possible. In the rear area, sufficient storage space is provided for additional luggage as well as the stowage of a folded standard wheelchair. The wheelchair is secured by a belt system in the back end of the cabin (FIG 28).

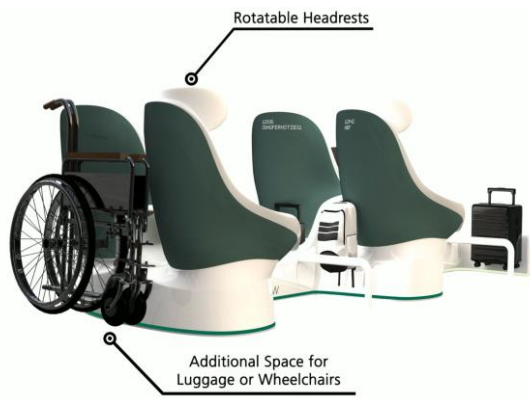

**FIG 28** Back/perspective view (Credits:DLR)

The central component of the overall concept is the seat. This is a lightweight seat consisting of a padded aluminum frame and a natural fiber fabric. Functionally, the principle was adopted from camping chairs, so that the upholstery fabric with light padding is already sufficient to support the back and buttocks.

In addition to the cabin design, an exemplary integration was integrated into a tiltrotor vehicle concept, which was developed in the course of an overall fleet development and system integration in the Horizon UAM project (FIG 29) [30].

## 3.5. Contribution and benefits for overall UAM system

Within the Horizon UAM project, various factors were considered in the examination and development of the overall UAM system. By applying the Design Thinking approach to the cabin design process for future air taxis, important advantages for the development process of the air taxi system as a whole were identified. The key findings are described below:

**Potential for weight reduction**

Initial estimates based on commercial aircraft interior masses suggest a possible total weight of 766 kg (for the Airport Shuttle concept) and 639 kg (for the Intracity concept). By applying DT and by using lightweight components, a minimalist seat design and a simplified luggage storage solution, the cabin mass could be reduced substantially to 380 kg to 579 kg. The weight reduction can have a positive impact on the sizing of the battery and the required power for air taxi transportation.

In addition, the user-centered design process enables further potential for cost savings. The early feedback process as well as the direct translation of main requirements into a concept can help to avoid cost-intensive adjustments in the late development process.

**Acceptance & Safety**

The direct involvement of different user groups in the design process offers great potential for creating awareness about UAM, disseminating information, and addressing concerns and fears. Particularly during a Focus Group Study, an increased acceptance of this novel mode of transportation was observed. The passengers influence on the design can have a positive impact on acceptance and the perception of safety, leading to a higher willingness to use air taxis among the general population. Moreover, by addressing fears, desires, and concerns directly and incorporating them into the design concept in collaboration with user groups, the development process of autonomously operated air taxis can lead to increased acceptance in the next step.

**Flexibility & Convenience**

By combining the Airportshuttle and Intracity scenario, a multifunctional cabin concept has been developed covering two different application scenarios. The versatility of the cabin concept leads to lower development costs compared to designing separate cabins for each use case. At the same time, the recognition value of the concept increases when used in multiple scenarios, which can positively impact the perceived safety and usability of the cabin features. With its minimalist and interchangeable cockpit design, the cabin can also be changed into a fully autonomous scenario with four passengers in the future.

In addition to the improved seating comfort, optimized storage compartments, minimalist design and customizable privacy features, the cabin design incorporates various comfort parameters based on the feedback from potential user groups. The deliberate combination of minimalist and easily understandable functions with futuristic and complex design elements enhances the overall comfort and user experience.

## 4. CONCLUSION AND OUTLOOK

This study demonstrated how future user groups can be actively involved in the cabin design process for future air taxis using the creative method of Design Thinking. The Design Thinking variant of the d.school was utilized and the process steps were tailored to the specific design process. In the initial process step (Empathize), different user groups were specifically involved from the beginning to gain an understanding of their experiences in public transportation and to identify their requirements and desires for the design of an air taxi cabin. Further engagement with user groups was conducted through an online study, where ideas based on the Focus Group Study and the identified priority areas were evaluated. Following the Empathize, Define, and Ideate process steps, a digital prototype was subsequently developed. This concept serves as a basis for the development of a complete vehicle and the related disciplines within the Horizon UAM project, such as integration into the overall system (Airportshuttle & Intracity concept), maintenance, battery system design, rotor concept development, and other areas. The next step focuses on the final process step, Assess. By employing immersive testing and simulation scenarios through mixed reality and physical mockups, different cabin designs and functional elements can be tested by user groups. An effective evaluation process can increase the maturity of the overall airtaxi concept and optimize acceptance investigations of various cabin scenarios. Initial studies have already been conducted within the project, indicating a high potential for further exploration and development of air taxi concepts [31] [32] [33].

## DECLARATIONS

### Competing interests

The authors have no competing interests to declare that are relevant to the content of this article.

## REFERENCES

[1]   Reuter, F.: "www.volocopter.com," 21 June 2021. https://www.volocopter.com/de/newsroom/volocopter-fliegt-ueber-paris/. [Accessed 10 July 2023].

[2]   Eißfeld, H.,Vogelpohl, V., Stolz, M., Papenfuß, A., Biella, M., Belz, J. and Kügler, D.: "The acceptance of Civil Drones in Germany," in *CEAS Aeronautical Journal*, Braunschweig, 2020.

[3]   End, A., Barzantny, C.,Stolz, M., Papenfuß, A. and Schmidt, R.: "Are you excited or worried about civilian drones? A large-scale telephone survey in the German population," in *3rd Horizon UAM Symposium*, Cochstedt (Germany), 2023.

[4]   EASA, Study on the societal acceptance of Urban Air Mobility in Europe, EU, 2021.

[5]   Cameron E. and Murray J.: www.pwc.co.uk," 2019. https://www.pwc.co.uk/issues/emerging-technologies/drones/drones-and-trust.html. Accessed 10 Juni 2023

[6]   Kantar - Department for Transport, "Transport and technology: Public Attitudes Tracker, Wave 4 summery report," Kantar Public, England, 2019.

[7]   Gaiser,   C.:   "www.bdli.de,"   24   June   2022. https://www.bdli.de/meldungen/akzeptanz-gegenueber-drohnen-und-flugtaxis-deutschland-steigt. Accessed 12 June 2023

[8]   Kellermann, R., Biehle T. Fischer, L.: Drones for parcel and passenger transportation: A literature review," in *Transportation Research Interdisciplinary Perspectives*, Elsevier, 2020.

[9]   Prof. Dr. Buhrmester, M.: www.uid.com, 3 March 2016. https://www.uid.com/de/aktuelles/hcd-design-thinking. Accessed 12 July 2023

[10]  Meinel, C., Leifer, L. and Plattner, H.: *Design thinking Research: Understanding- improve- apply. Understanding Innovation*, Heidelberg, Springer, 2011, p. 14.

[11]  Gruber, M., de Leon, N., Geroge G. and Thompson, T.:   Managing   by   design:   From   the editors.Acad.Manag.J.,   The   Academy   of Management Journal, 2015, pp. 58,1-7.

[12]  Steward, S., Giambalvo, J., Vance, J., Faludi J. and Hoffenson S.: A Product Development Approach Advisor for Navigating Common Design Methods, Processes, and Environments," in *Designs 2020*, 2020, pp. 4,4..

[13]  Kwon, J., Choi, Y. and Hwang,Y.: Enterprise Design Thinking: An Investigation on User-Centered Design Processes in Large Corporations, Designs, College of Design, University of Minnesota Twin-Cities, Minneapolis, USA, 2021.

[14]  www.heliosdesign.com: How to apply the principles of IBM Enterprise Design Thinking to your next project. https://www.heliosdesign.com/blog/web/how-to-apply-ibm-enterprise-design-thinking-principles.html. Accessed 2 July 2023

[15]  www.ibm.com: The Loop drives us Understand the present and envision the future in a continuous cycle of observing, reflecting, and making. https://www.ibm.com/design/thinking/page/framework/loop. Accessed 12 July 2023

[16]  Choudhary, S.: Design Thinking: Divergence and Convergence   Cycles. https://medium.com/@i.shubhangich/design-thinking-divergence-and-convergence-cycles-3ce7a6f27815. Accessed 12 July 2023

[17]  Herrmann, D.: Design Thinking Double Diamond - 2 Diamanten   für   brilliantes   Design? https://betasphere.de/de/design-thinking-double-diamond. Accessed 12 July 2023

[18]  Balcaitis, R.: Design Thinking models. Stanford d.school.   https://empathizeit.com/design-thinking-models-stanford-d-school/. Accessed 10 July 2023

[19] Doorley, S., Holcomb, S., Klebahn, P., Segovia, K., and Utley, J.: *design thinking bootleg,* Stanford, USA: d.school Hasso Plattner Institute of Design at Stanford University, 2018.

[20] Dr. Schuchardt, B.-I.: *Horizon UAM Project Overview,* Urban Air Mobility Virtial Symposium: German Aerospace Center e.V. (DLR), 2021.

[21] Asmer, L., Pak, H., Prakasha, P., S., Dr.Schuchardt, B.-I., Weiand, P., Meller, F., Torens, C., Becker, D., Zhu, C., Schweiger, K., Volkert, A. and Jaksche,R.: *Urban Air Mobility Use Cases and Technology Scenarios for the Horizon UAM Project,* Urban Air Mobility Virtual Symposium: German Aerospace Center e.V. (DLR), 2018.

[22] Moerland-Masic, I., Reimer, F. and Bock, T.-M.: "Urban Mobility: Airtaxi Cabin from a Passengers Point of View," in *International Comfort Congress* , Nottingham, UK, 2021.

[23] www-aerospace-technology.com: Lilium Seven-Seater eVTOL Jet, Germany , https://www.aerospace-technology.com/projects/lilium-7-seater-evtol-jet/attachment/image-4-lilium-7-seater-evtol-jet/. Accessed 12 July 2023

[24] www.nahverkehrspraxis.de: MOIA baut Ridepooling-Service in Hamburg aus. https://www.nahverkehrspraxis.de/moia-baut-ridepooling-service-in-hamburg-aus. Accessed 12 July 2023

[25] Stolz M., Laudien T., Reimer F. and Moerland-Masic, I.: A USer-Centered Cabin Design Approach to Investigate Peoples Preferences on the Interior Design of Future Air Taxis in *DASC 2021,* The 40th Digital Avionics Systems Conference (virtual), 2021.

[26] Brown S.: *Likert scale examples for surveys,* Iowa State University, USA: ANR Program evaluation, 2010.

[27] Zec, M.: Walt-Disney-Methode. https://xn--kreativittstechniken-jzb.info/ideen-generieren/walt-disney-methode/. Accessed 12 July 2023

[28] Eich, D.J.: How to Create Personas for Design Thinking. https://www.innovationtraining.org/create-personas-design-thinking/. Accessed 12 June 2023

[29] Reimer, F., Masic, I.-M., End, A., Schadow J., Bock, T.-M., Meller, F. and Nagel, B. "Safety & Privacy in Urban Air Mobility (UAM) - A User Centric Design Approach Providing Insights into People´s Preferences for UAM Cabin Designs," in *AHFE Conference 2022,* New York City, USA, 2022.

[30] Ratei, P., Naeem, N. P. Prakasha, P.-S.: Development of an Urban Air Mobility Vehicle Family Concept by System of Systems Aircraft Design and Assessment in *12th EASN International Conference, 18-21 Oct 2022,* Barcelona, Spain, 2022.

[31] Laudien, T., Ernst, J.M. and Schuchardt, B.I.: Implementing a Customizable Air Taxi Simulator with a Video-See-Through Head-Mounted Display–A Comparison of Different Mixed Reality Approaches in *IEEE/AIAA 41st Digital Avionics Systems Conference (DASC),* Portsmouth, VA, USA, 2022.

[32] Laudien, T., Papenfuß, A., Ernst, J.M., Stolz, M., Schuchardt, B.I. and End, A.: Assessment of Air Taxi Passenger Acceptance– Implementation and Initial Evaluationof a Mixed Reality Simulator in *2nd Urban Air Mobility Symposium*, Braunschweig, Germany, 2022.

[33] Ernst, J.M., Laudien, T., Lenz, H. and Schuchardt, B.I.:Entwicklung eines konfigurierbaren Flugtaxi-Simulators mit Hilfe einer kopfgetragenen Anzeige mit Video-Durchsicht-Ein Vergleich verschiedener Mixed-Reality-Ansätze in *Deutscher Luft und Raumfahrt Kongress 2022*, Dreseden, Germany, 2022.